\documentclass[aps,prstab,twocolumn,superscriptaddress]{revtex4-2}

\usepackage{graphicx}
\usepackage{booktabs}
\usepackage{amssymb}
\usepackage{graphics}
\usepackage{amsmath}
\usepackage{placeins}
\usepackage{hyperref}
\usepackage{verbatim}
\usepackage{lineno,hyperref}
\usepackage[dvipsnames]{xcolor}

\begin{document}


\title{Eliminating uncertainty of thermal emittance measurement in solenoid scans due to rf and solenoid fields overlap}


\author{Lianmin Zheng}
\email[]{zhenglm14@tsinghua.org.cn}
\affiliation{Department of Engineering Physics, Tsinghua University, Beijing 100084, People's Republic of China}
\affiliation{Key Laboratory of Particle and Radiation Imaging, Tsinghua University, Ministry of Education, Beijing 100084, People's Republic of China}
\author{Yingchao Du}
\affiliation{Department of Engineering Physics, Tsinghua University, Beijing 100084, People's Republic of China}
\affiliation{Key Laboratory of Particle and Radiation Imaging, Tsinghua University, Ministry of Education, Beijing 100084, People's Republic of China}

\author{Pengwei Huang}
\affiliation{Department of Engineering Physics, Tsinghua University, Beijing 100084, People's Republic of China}
\affiliation{Key Laboratory of Particle and Radiation Imaging, Tsinghua University, Ministry of Education, Beijing 100084, People's Republic of China}

\date{\today}

\begin{abstract}
The solenoid scan is one of the most common methods for the in-situ measurement of the thermal emittance of a photocathode in an rf photoinjector. The fringe field of the solenoid overlaps with the gun rf field in quite a number of photoinjectors, which makes accurate knowledge of the transfer matrix challenging, thus increases the measurement uncertainty of the thermal emittance. This paper summarizes two methods that have been used to solve the overlap issue and explains their deficiencies. Furthermore, we provide a new method to eliminate the measurement error due to the overlap issue in solenoid scans. The new method is systematically demonstrated using theoretical derivations, beam dynamics simulations, and experimental data based on the photoinjector configurations from three different groups, proving that the measurement error with the new method is very small and can be ignored in most of the photoinjector configurations.
\end{abstract}


\maketitle

\section{Introduction}
Thermal emittance, i.e., the mean transverse momentum of the electrons emitted from a cathode, is an extremely significant figure of merit for photoinjectors because the transverse emittance can be efficiently preserved during transmission in modern linear accelerators thus the thermal emittance heavily determines the final emittance at the end of a photoinjector. Therefore, intense studies have focused on the thermal emittance characterizations, and the efforts of the thermal emittance diagnosis and minimization will finally benefit many photoinjector-based machines for scientific research \cite{emma2010first,weathersby2015mega,du2013generation}.

\par The solenoid scan is a widely used method for the in-situ measurement of the thermal emittance in a photoinjector \cite{bazarov2008thermal,hauri2010intrinsic,qian2012experimental,lee2015review,maxson2017direct,graves2001measurement,gulliford2013demonstration,miltchev2005measurements,bazarov2011thermal}. The measurement adopts a simple experimental setup: a dc or rf photocathode gun successively followed by a solenoid and a drift. The photoemission electron beam is accelerated to relatively high energy by the photogun and then focused by the solenoid onto a fluorescent screen located at the end of the drift. The thermal emittance of the photocathode can be obtained by fitting the measured beam spot sizes on the screen as a function of the solenoid strength.

In recent years numerous efforts have been made to improve the accuracy of the thermal emittance measurement via the solenoid scan method, and these efforts can be divided into three categories. The first one is the reduction of the emittance growth in beam transmission. There are several factors that increase the beam emittance thus leading to an overestimation of the thermal emittance, such as (i) nonlinear space charge (SC) effects \cite{lee2015review,miltchev2005measurements}, (ii)  rf effects in the gun \cite{chae2011emittance}, (iii) spherical, chromatic and coupled transverse dynamics aberrations in the solenoid \cite{dowell2016sources, mcdonald1989methods,dowell2018exact,zheng2018overestimation,PhysRevAccelBeams.22.072805}. SC effects can be alleviated by using low charge beams while rf effects can be mitigated by using short beams. The spherical and coupled transverse dynamics aberrations can be reduced by keeping the beam size small inside the solenoid. The chromatic aberration can be mitigated by operating with a short bunch at low charge \cite{dowell2016sources}. In conclusion, these factors have been well studied and the overestimation of the thermal emittance due to these factors can be eliminated.

The second category of reducing the error of the thermal emittance measurement is the accurate
beam size measurement on the screen. Since low charge beam is required to reduce the emittance growth due to SC effects in beam transmission, the electron beam images on the fluorescent screen are usually dim, and accurate beam size calculation based on these images is challenging. Fortunately, the image quality can now be well improved by employing high sensitivity CCD cameras \cite{qian2012experimental} and thin YAG:Ce screens with high resolution \cite{maxson2017direct}. 

Finally, the third category requires accurate knowledge of the transfer matrix. Only the elements between the solenoid entrance and the screen are involved in the transfer matrix calculation for an idealized solenoid scan configuration. Therefore, the transfer matrix can be calculated accurately as long as the solenoid field profile, the solenoid peak strength, and the drift length between the solenoid and the screen are accurately measured. However, the realistic beamline configuration of a photoinjector is much more complicated. The solenoid's axial field profile is not ideally hard-edged, and the fringe field of the solenoid usually overlaps with the gun rf field in quite a number of photoinjectors, especially in normal-conducting photoinjectors.  In this case, the accurate knowledge of the transfer matrix becomes pretty challenging.

Numerous results of thermal emittance measurement in rf photoinjectors based on the solenoid scan technique have been published in recent decades. Probably most of the previous works focused on the physics behind the results of thermal emittance measurement instead of the solenoid scan technique itself, the efforts to solve the rf and solenoid fields overlap issue were only briefly mentioned or omitted altogether. To our knowledge, some methods have been conventionally employed in the previous solenoid scan works to solve the overlap issue. These methods, however, usually increase the measurement uncertainty or even lead to an overestimation of the thermal emittance.  We believe it is necessary to summarize these methods and point out their deficiencies, and more importantly, this paper aims to provide a new method to eliminate the uncertainty of the thermal emittance measurement in solenoid scans due to rf and solenoid fields overlap.

This paper is organized as follows. Sec.~\ref{section2} briefly describes the solenoid scan formalism without the fields overlap issue. Sec.~\ref{section3} offers three fields overlap examples in different photoinjector beamlines. Sec.~\ref{section4} summarizes two conventionally used methods to solve the overlap issue and their deficiencies. Sec.~\ref{section5} theoretically provides a new method to solve the overlap issue, which can eliminate the uncertainty of the thermal emittance measurement in solenoid scans. Finally, beam dynamics simulations and experiments are demonstrated in Sec.~\ref{section6} and Sec.~\ref{section7} respectively to verify the performance of the new method compared with the previously used methods.

\section{solenoid scan formalism without the overlap issue}\label{section2}

\par Firstly we demonstrate the basic formalism of the solenoid scan without the fields overlap issue. Only a transport line consisting of a solenoid and a drift is involved in the transfer matrix calculation.  Assuming the solenoid is hard-edged and the axial magnetic field inside the solenoid is  $B_0$, the length of the solenoid is $L$, and the length of the drift is  $L_d$. 

The solenoid's transfer matrix can be expressed as
\begin{equation}\label{sol_matrix}
{R_{sol}} = \left[ {\begin{array}{*{20}{c}}
	{{C^2}}&{\frac{{SC}}{K}}&{SC}&{\frac{{{S^2}}}{K}}\\
	{ - KSC}&{{C^2}}&{ - K{S^2}}&{SC}\\
	{ - SC}&{ - \frac{{{S^2}}}{K}}&{{C^2}}&{\frac{{SC}}{K}}\\
	{K{S^2}}&{ - SC}&{ - KSC}&{{C^2}}
	\end{array}} \right]=R_{rot}R_{foc}
\end{equation}
where $S\equiv{\rm sin}(KL)$, $C\equiv{\rm cos}(KL)$. $K \equiv (eB_0)/(2p)$, $KL$ denote strength, and Larmor angle of the solenoid, respectively; $e$ and $p$ denote charge and momentum of electron. $R_{rot}$ and $R_{foc}$ are the rotation matrix and the focusing matrix respectively.
\begin{equation}
{R_{rot}} = \left[ {\begin{array}{*{20}{c}}
C&0&S&0\\
0&C&0&S\\
{ - S}&0&C&0\\
0&{ - S}&0&C
\end{array}} \right]
\end{equation}
\begin{equation}
{R_{foc}} = \left[ {\begin{array}{*{20}{c}}
C&{\frac{S}{K}}&0&0\\
{ - KS}&C&0&0\\
0&0&C&{\frac{S}{K}}\\
0&0&{ - KS}&C
\end{array}} \right]
\end{equation}

The drift's matrix can be expressed as
\begin{equation}\label{R_d}
{R_d} = \left[ {\begin{array}{*{20}{c}}
	1&{{L_d}}&0&0\\
	0&1&0&0\\
	0&0&1&{{L_d}}\\
	0&0&0&1
	\end{array}} \right]
\end{equation}
where $L_{d}$ is the length of the drift.

The thermal emittance $\epsilon_x$ and $\epsilon_y$ are complicated to deduce because the beam trajectories in x and y directions are coupled due to the rotation matrix of the solenoid. For simplicity the rotation term $R_{rot}$ is usually ignored in the beam moments calculation \cite{qian2012experimental,lee2015review,Scifo2018}, and the transfer matrix of the solenoid scan beamline is expressed as $R\equiv R_d R_{foc}$.

The beam spot size squared taken on the screen at the end of the drift can be expressed as the function of the transfer matrix:
	\begin{equation}\label{eq1}
	\sigma ^2=R_{11}^{2}\left\langle x_0^2 \right\rangle
	+2R_{11}R_{12}\left\langle {x_0}{x_0}^\prime \right\rangle
	+R_{12}^{2}\left\langle x_{0}^{\prime 2} \right\rangle,
	\end{equation}
	where $\left\langle x_0^2 \right\rangle $, $\left\langle {x_0}{x_0}^\prime \right\rangle $ and $\left\langle x_{0}^{\prime 2} \right\rangle $ are the beam moments at the solenoid entrance. $R_{11}$ and $R_{12}$ are the elements of the transfer matrix $R$.

Therefore, the beam moments can be fitted based on the beam size varying with the solenoid strength, and the normalized emittance at the solenoid entrance can be written as 
	\begin{equation}\label{eq2}
	{\varepsilon _n} = \beta \gamma \sqrt {\left\langle {x_0^2} \right\rangle \left\langle x_{0}^{\prime 2} \right\rangle  - {{\left\langle {{x_0}{x_0}^\prime } \right\rangle }^2}} 
	\end{equation}

\section{rf and solenoid fields overlap}\label{section3}
\begin{figure}[hbtp]
	\centering
	\includegraphics[scale=0.8]{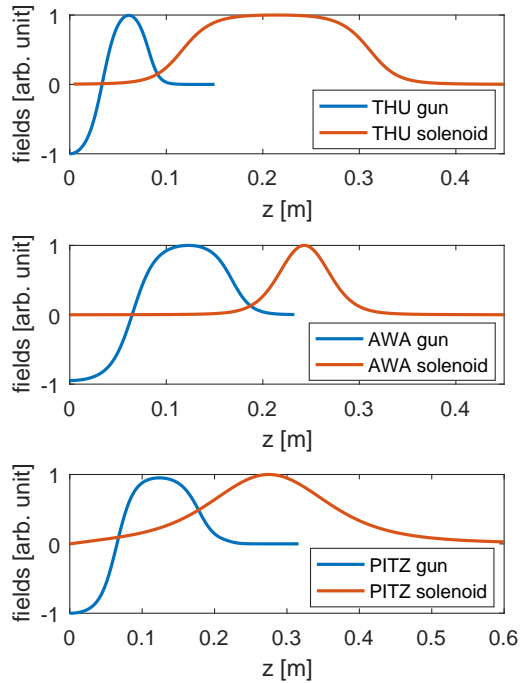}
	\caption{\label{Fig.field_plot}  rf and solenoid field profiles of the  THU, AWA and PITZ photoinjectors. The blue lines indicate the rf field and the red lines indicate the solenoid field. Both the rf and the solenoid field profiles are normalized by the respective peak magnitude. }
\end{figure}

Different from the ideal case demonstrated in Sec.~\ref{section2}, the real solenoid is not hard-edged and the fringe field overlaps with the rf field in most of the normal-conducting photoinjectors. For example, Fig.~\ref{Fig.field_plot} depicts the profile of the axial rf and solenoid fields of the photoinjectors from three different groups: Accelerator Laboratory of Tsinghua University (THU), Argonne Wakefield Accelerator (AWA), and Photo Injector Test Facility at DESY Zeuthen (PITZ). The frequency is 2.856 GHz for the THU gun, and is 1.3 GHz for the AWA and DESY guns. The rf and solenoid fields overlap in all photoinjector configurations. The PITZ gun has a bucking solenoid to eliminate the axial field of the main solenoid at the cathode so that the main solenoid can be placed closer to the gun. Therefore, the rf and solenoid fields of the PITZ photoinjector overlap more than the AWA and THU photoinjectors. 

\begin{figure*}[hbtp]
	\centering
	\includegraphics[scale=0.8]{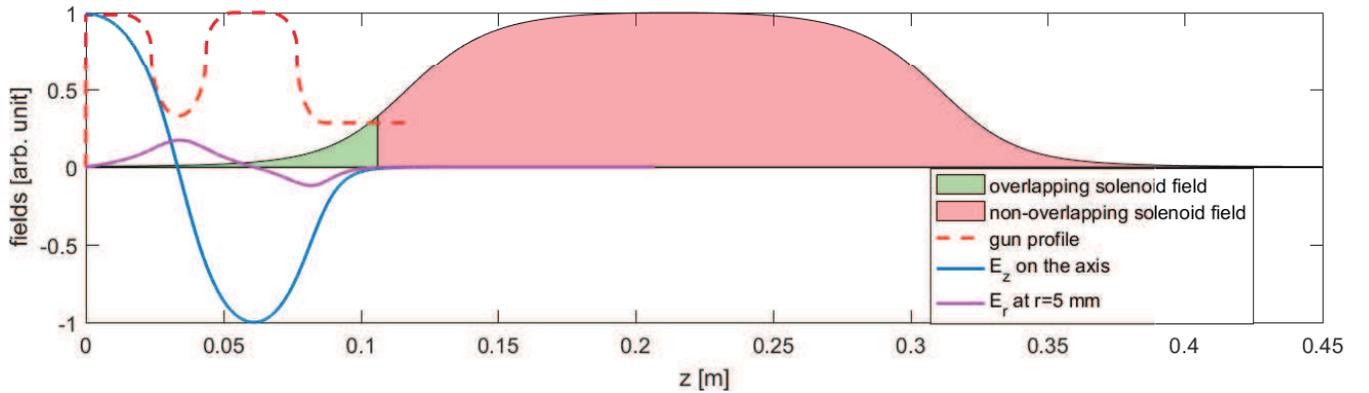}
	\caption{\label{Fig.Tsinghua_field_distribution} Field distributions of the THU photoinjector. The dashed red line exhibits the gun profile. The blue line indicates the longitudinal electric field on the axis and the magenta line indicates the radial electric field at an off-axis position $r$=5 mm. The light green area indicates that the axial solenoid field overlaps with the rf field, and the light red area indicates that the field that doesn't overlap with the rf field.}
\end{figure*}

A more detailed illustration of the fields overlap, taking the THU photoinjector as an example, is depicted in Fig.~\ref{Fig.Tsinghua_field_distribution}. The axial electric field at the gun exit gradually decreases along the z-axis. We define a position $z=L_{sep}$ where 
 
\begin{equation}
E_z(L_{sep})=E_0\times1\%
\end{equation}

 The rf field can be ignored where $z>L_{sep}$ since $E_z(z>L_{sep})$ is less than the one-hundredth of the maximum axial electric field. $L_{sep}$ is a boundary that divides the solenoid into two parts: the left part (light green area in Fig.~\ref{Fig.Tsinghua_field_distribution}) overlaps with the rf field, and the right part (light red area) does not. 
 
The rf gun defocuses the beam transversely. The radial electric field of the THU gun (present in Fig.~\ref{Fig.Tsinghua_field_distribution}) is largest at the center iris and the exit iris, indicating that the transverse force is largest at these two irises.  Kim \cite{kim1989rf} and Dowell \cite{rao2014engineering}'s theory proves that the transverse force at the center iris is negligible since $E_z(z)$ is anti-symmetric about the iris for $\pi$-mode field. Thus, only the transverse force at the exit iris is significant, and the total transverse force is an impulse given at the exit iris, which can be simplified into a transverse defocusing thin lens with a focal length of 
\begin{equation}
{f_{g}} =  - \frac{{2\beta \gamma m{c^2}}}{{e{E_0}\sin {\phi _e}}}
\end{equation}
where $e$, $m$, $c$ describe the elementary charge, the electron rest mass, and the speed of light, respectively. $\beta$ and $ \gamma$ are the Lorentz factors. $E_0$ and $\phi _e$ are the peak axial electric field and the rf phase when the beam arrives at the gun exit, respectively.

The rf gun accelerates the beam longitudinally. The geometric phase space  $\left[ {\begin{array}{*{20}{c}}
x\\
{x'}
\end{array}} \right]$ changes in an accelerating field since the angle of divergence $x'=p_x/p_z$ reduces with increasing $p_z$. Therefore, the geometric phase space can not be directly employed in the transfer matrix calculation. Some work \cite{peiliang2021simulation} used normalized phase space  $\left[ {\begin{array}{*{20}{c}}
x\\
{p_x}
\end{array}} \right]$ to calculate the transfer matrix in an rf gun. In this work we choose another way. We assume that the beam momentum from the cathode to the screen is a constant equal to $p_f$, where $p_f$ is the beam momentum after the gun exit. Based on this assumption the geometric phase space can be used since $x'$ is conserved in the beam transport. The field of the overlapping solenoid is replaced by a momentum-related equivalent field
\begin{equation}\label{eq1009}
B_z'\left( z \right) = B_z\left( z \right) \times {p_f}/p_z(z)
\end{equation}
to ensure that the transverse beam dynamics in the rf gun remain unchanged. Here $B_z(z)$ is the real field of the overlapping solenoid, $p_z(z)$ is the real beam momentum at position $z$.  In
order to keep the analysis simple, the geometric phase space is used in the transfer matrix calculation in the following derivations based on the above transformation.

  Besides, the ratio of the integral strength of the two parts of the solenoid $\xi$ is defined as:
\begin{equation}\label{eq1008}
\xi {\rm{ = }}\frac{{\int_0^{{{L}_{sep}}} {{B'_z}^2(z)dz} }}{{\int_{{{L}_{sep}}}^{ + \infty } {{B_z}^2(z)dz} }}
\end{equation}

$\xi$ is an important value that determines the  overlap severity in a solenoid scan beamline, and will be used in the following analytic expressions and numerical simulations. Note that $\xi$ is a momentum-related quantity because $B'_z(z)$ contains the term of $p_z(z)$.

\begin{figure}[hbtp]
	\centering
	\includegraphics[scale=0.42]{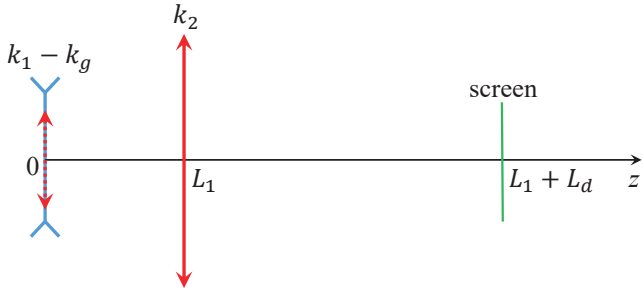}
	\caption{\label{Fig.sketch} A simplified model of the overlapped rf and solenoid fields in solenoid scan beamline.}
\end{figure}

Based on the analysis above, a simplified model of the overlapped rf and the solenoid fields is built, as  illustrated in Fig.~\ref{Fig.sketch}. Position 0 is located at the gun exit. The rf field is simplified into a thin defocusing lens at the gun exit with the defocusing strength of $-k_g$. The solenoid field is simplified into two thin focusing lenses with the focusing strength of $k_1$ and $k_2$ respectively. $k_1$ represents the overlapping solenoid and $k_2$ represents the non-overlapping solenoid. Note that the position of $k_1$ is not strictly located at the gun exit in the thin-lens approximation. However, considering that both the defocusing strength of $-k_g$ and the focusing strength of $k_1$ are weak, i.e., the beam size from the cathode to the gun exit is roughly a constant, the position of $k_1$ moving a little to the gun exit will not change the beam dynamics. Therefore, we assume that $k_1$ overlaps with $-k_g$ in Fig.~\ref{Fig.sketch}. The distance between the two parts of the solenoid field ($k_1$ and $k_2$) is assumed to be $L_1$. The distance between the right part of the solenoid field ($k_2$) and the screen is assumed to be $L_d$. The overlapping solenoid field strength is proportional to the non-overlapping solenoid field strength in solenoid scans. Based on Eqn.~\ref{eq1008},  the relation of $k_1$ and $k_2$ can be expressed as ${k_1} = \xi {k_2}$.

\section{ conventionally used methods to solve the fields overlap issue and their deficiencies}\label{section4}
As far as we know, there are two main methods that have been used in previous solenoid scan works to solve the rf and solenoid fields overlap issue. $\it Method~One$ is a commonly used method that can avoid the knowledge of the rf field. It calculates the transfer matrix in the following way:  the overlapping solenoid field is abandoned and the transfer matrix of the non-overlapping solenoid field is involved in thermal emittance fitting, thus the formalism introduced in Sec.~\ref{section2} can be employed without theoretical problem. The deficiency of this method is that it leads to an overestimation of the thermal emittance. It should be emphasized that this overestimation is not due to the beam's azimuthal momentum obtained from the rotation motion in the overlapping solenoid, because the rotation matrix $R_{rot}$ has been ignored in the thermal emittance fitting,  as introduced in Sec.~\ref{section2}, and there is no x-y dimension coupling in the theory of the solenoid scan.  The real source of the overestimation is that the effect of the varying field strength of the overlapping solenoid on the change of the beam spot size at the screen is ignored.

 The overestimation of the thermal emittance in the $\it Method~One$ is analyzed based on transfer matrix calculations with the simplified beamline demonstrated in Fig.~\ref{Fig.sketch}. As illustrated in Sec.~\ref{section3}, the geometric phase space is used in transfer matrix calculation. The initial beam is characterized by the beam sigma matrix at position 0 (before $k_1$ and $-k_g$). In order to keep the analysis simple, we assume the initial beam has zero-emittance, uniform transverse distribution, with the same beam size squared in both x and y directions ($\langle {x_0}^2\rangle$) and perfectly parallel rays. Since there is no x-y dimension coupling in the beamline, two-dimensional transfer matrices are employed in the following calculations. The initial beam matrix and initial emittance at position 0 can be expressed as
\begin{equation}\label{eq150}
\begin{array}{l}
{\Sigma _0} = \left[ {\begin{array}{*{20}{c}}
	{\langle {x_0}^2\rangle}&0\\
	0&0\\
	\end{array}} \right] \\
 \varepsilon_{0}=0
\end{array}
\end{equation}
Eqn.~\ref{eq150} completely specifies the initial beam conditions. 

The transfer matrix of the rf field can be expressed as  ${R_g} = \left[ {\begin{array}{*{20}{c}}
1&0\\
k_g&1
\end{array}} \right]$, where ${k_g} =- \frac{1}{f_g}$. Similarly, the transfer matrix of the overlapping and non-overlapping solenoid can be expressed as ${R_{s1}} = \left[ {\begin{array}{*{20}{c}}
1&0\\
{ - {k_1}}&1
\end{array}} \right]$, ${R_{s2}} = \left[ {\begin{array}{*{20}{c}}
1&0\\
{ - {k_2}}&1
\end{array}} \right]$, respectively. Therefore, the total transfer matrix from the position 0 (before $k_1$ and $-k_g$) to the screen can be expressed as 
\begin{equation}\label{eq9}
R = \left[ {\begin{array}{*{20}{c}}
1&{{L_d}}\\
0&1
\end{array}} \right]\left[ {\begin{array}{*{20}{c}}
1&0\\
{ - {k_2}}&1
\end{array}} \right]\left[ {\begin{array}{*{20}{c}}
1&{{L_1}}\\
0&1
\end{array}} \right]\left[ {\begin{array}{*{20}{c}}
1&0\\
{{k_g}}&1
\end{array}} \right]\left[ {\begin{array}{*{20}{c}}
1&0\\
{ - {k_1}}&1
\end{array}} \right]
\end{equation}

Here ${k_1} = \xi {k_2}$, where $\xi$ is the strength ratio of the two parts of the solenoid defined in Sec.~\ref{section3}. The beam sigma matrix at the screen can be expressed as $\Sigma  = R{\Sigma _0}{R^T}$, so that the beam size squared at the screen 
as a function of the solenoid strength $k_2$ can be expressed as

\begin{equation}\label{eq10}
\begin{array}{l}
{\sigma _{{scr}}}^2 = \Sigma (1,1) = \left\langle {{x_0}^2} \right\rangle  \times \\
{[1 - {k_2}{L_d} - (\xi {k_2} - {k_g})({L_1} + {L_d} - {k_2}{L_1}{L_d})]^2}
\end{array}
\end{equation}

Based on the $\it Method~One$, the transfer matrix of the solenoid scan beamline should be $M = \left[ {\begin{array}{*{20}{c}}
1&{{L_d}}\\
0&1
\end{array}} \right]\left[ {\begin{array}{*{20}{c}}
1&0\\
{ - {k_2}}&1
\end{array}} \right]$. To retrieve the emittance measured by the solenoid scan, the beam sizes squared at the screen ${\sigma _{{scr}}}^2$ are fitted based on the transfer matrix $M$ in order to obtain the fitted beam moments before $k_2$: $\left\langle {{x_{fit}}^2} \right\rangle $, $\left\langle {{x_{fit}}{x_{fit}}^\prime } \right\rangle $ and $\left\langle {{x_{fit}}{{^\prime }^2}} \right\rangle $. The fitted beam sizes squared ${\sigma _{fit}}^2$ as a function of the fitted beam moments can be expressed as

\begin{equation}
{\sigma _{fit}}^2 = {M_{11}}^2\left\langle {{x_{fit}}^2} \right\rangle  + 2{M_{11}}{M_{12}}\left\langle {{x_{fit}}{x_{fit}}^\prime } \right\rangle  + {M_{12}}^2\left\langle {{x_{fit}}{{^\prime }^2}} \right\rangle 
\end{equation}

 We calculate ${\sigma _{scr}}^2 - {\sigma _{fit}}^2$ and factorize it in terms of $k_2$:

\begin{equation}
{\sigma _{scr}}^2 - {\sigma _{fit}}^2 = {p_0} + {p_1}{k_2} + {p_2}{k_2}^2 + {p_3}{k_2}^3 + {p_4}{k_2}^4
\end{equation}

The expressions of the coefficient $p_0$, $p_1$ and $p_2$ are too long to be written here, but they contain the terms of beam moments $\left\langle {{x_{fit}}^2} \right\rangle $, $\left\langle {{x_{fit}}{x_{fit}}^\prime } \right\rangle $ and $\left\langle {{x_{fit}}{{^\prime }^2}} \right\rangle $.

The coefficients of the term ${k_2}^3$ and ${k_2}^4$ can be expressed as

\begin{equation}
{p_3} =  - 2\xi {L_1}{L_d}\left\langle {{x_0}^2} \right\rangle \left( {{L_d} + {k_g}{L_1}{L_d} + \xi {L_1} + \xi {L_d}} \right)
\end{equation}
\begin{equation}
{p_4} = {\xi ^2}{L_1}^2{L_d}^2\left\langle {{x_0}^2} \right\rangle 
\end{equation}

We found that both ${p_3}$ and ${p_4}$ don't contain the terms of beam moments $\left\langle {{x_{fit}}^2} \right\rangle $, $\left\langle {{x_{fit}}{x_{fit}}^\prime } \right\rangle $ and $\left\langle {{x_{fit}}{{^\prime }^2}} \right\rangle $. Moreover, ${p_3} \ne 0$ and ${p_4} \ne 0$ if the rf and solenoid fields overlap ($\xi \ne 0$). Therefore, there is no any certain $\left\langle {{x_{fit}}^2} \right\rangle $, $\left\langle {{x_{fit}}{x_{fit}}^\prime } \right\rangle $ and $\left\langle {{x_{fit}}{{^\prime }^2}} \right\rangle $ that make ${\sigma _{fit}}^2 = {\sigma _{scr}}^2$ under any solenoid strength $k_2$. The fitting routine of the $\it Method~One$ is to minimize $\left| {{\sigma _{scr}}^2 - {\sigma _{fit}}^2} \right|$ to retrieve the fitted beam moments $\left\langle {{x_{fit}}^2} \right\rangle $, $\left\langle {{x_{fit}}{x_{fit}}^\prime } \right\rangle $ and $\left\langle {{x_{fit}}{{^\prime }^2}} \right\rangle $.  The fitting error of $\left\langle {{x_{fit}}^2} \right\rangle $, $\left\langle {{x_{fit}}{x_{fit}}^\prime } \right\rangle $ and $\left\langle {{x_{fit}}{{^\prime }^2}} \right\rangle $ increases with the increase of $\xi$.

The fitted emittance can be expressed as 

\begin{equation}
{\varepsilon _{fit}} = \sqrt {\left\langle {{x_{fit}}^2} \right\rangle \left\langle {x{{_{fit}'}^2}} \right\rangle  - {{\left\langle {{x_{fit}}x_{fit}'} \right\rangle }^2}} 
\end{equation}

${\varepsilon _{fit}}$ is usually not equal to 0, thus ${\varepsilon _{fit}}$ can be regarded as the overestimation of the thermal emittance with the $\it Method~one$.

$\it Method~Two$ uses the beam dynamics simulation to replace the transfer matrix calculation to obtain the thermal emittance \cite{huang2020single}. A start-to-end simulation from the cathode to the screen is performed using a beam dynamics simulation code like ASTRA \cite{floettmann2011astra}, OPAL \cite{adelmann2019opal} or GPT \cite{de1996general}, etc. The information of the elements from the cathode to the screen, such as the field profiles, the peak fields of the gun and the solenoid, the length of the drift, and the position of these elements are imported into the code. An electron beam including a large number of macro-particles is generated with an assumed thermal emittance. The beam spot size at the screen varying with the solenoid strength is calculated in the simulation. The assumed thermal emittance is changed until the simulated spot sizes at the screen are in good agreement with the measurements, then the simulated thermal emittance is considered to be the real thermal emittance.  The limitation of this method is that it requires accurate knowledge of the beamline elements including the rf gun profile and the distance between the gun and the solenoid. The field profile of an rf gun is usually measured by the bead-pull method \cite{staufenbiel2005field,vlieks2002development}, and the measurement accuracy is too limited to be directly used in the beam dynamics simulation. As a compromise, the simulated rf field profile from CST or SUPERFISH is usually employed in the simulation. The thermal emittance measured in the $\it Method~Two$ will have unknown uncertainties if the rf field profile or the relative distance between the gun and the solenoid in the simulation has a discrepancy with the real beamline. Furthermore, this method can not even be used if the discrepancy is large.

\section{A new method to eliminate the uncertainty of thermal emittance measurement in solenoid scans}\label{section5}
A new method is provided to eliminate the uncertainty of thermal emittance measurement in solenoid scans due to the rf and solenoid fields overlap. The simplified beamline shown in Fig.~\ref{Fig.sketch} is still used in the following theoretical derivations, the same as the derivations demonstrated for the $\it Method~One$. The initial beam matrix and initial emittance are still expressed as the form shown in Eqn.~\ref{eq150}, and the initial emittance is zero.

The complete transfer matrix $R$ (Eq.~\ref{eq9}) including the rf field is difficult to calculate since the defocusing strength of the rf gun $k_g$ is unknown. We propose an equivalent transfer matrix $\tilde R$ to replace $R$ in the thermal emittance fitting, and $\tilde R$ is expressed as

\begin{equation}
\tilde R = \left[ {\begin{array}{*{20}{c}}
1&{{L_d}}\\
0&1
\end{array}} \right]\left[ {\begin{array}{*{20}{c}}
1&0\\
{ - {k_2}}&1
\end{array}} \right]\left[ {\begin{array}{*{20}{c}}
1&{{L_1}}\\
0&1
\end{array}} \right]\left[ {\begin{array}{*{20}{c}}
1&0\\
{ - {k_1}}&1
\end{array}} \right]
\end{equation}

Compared with $R$, $\tilde R$ only considers the complete solenoid field but not the rf field from position 0 to the screen. The last two terms of $R$ are $R_{s1}$ and $R_g$ respectively. Because the rf ($-k_g$) and solenoid ($k_{s1}$) fields overlap in the thin-lens approximation, and the phase advances of $R_{s1}$ and $R_g$ are zero, we have $R_g R_{s1}=R_{s1} R_g$. Therefore, the transfer matrix from the cathode to the screen can be expressed as 
\begin{equation}
R=\tilde R R_g
\end{equation}

As a result, the beam sigma matrix at the screen is
\begin{equation}
\Sigma  = R{\Sigma _0}{R^T}{\rm{ = }}\tilde R{R_g}{\Sigma _0}{(\tilde R{R_g})^T} = \tilde R({R_g}{\Sigma _0}{R_g}^T){{\tilde R}^T}
\end{equation}

$\Sigma _0$ is the real initial beam matrix on the cathode. If the equivalent transfer matrix $\tilde R$ is used in the thermal emittance fitting, the equivalent initial beam matrix on the cathode should be

\begin{equation}\label{eq2009}
{{\tilde \Sigma }_0} = {R_g}{\Sigma _0}{R_g}^T = \left[ {\begin{array}{*{20}{c}}
{\left\langle {{x_0}^2} \right\rangle }&{{k_g}\left\langle {{x_0}^2} \right\rangle }\\
{{k_g}\left\langle {{x_0}^2} \right\rangle }&{{k_g}^2\left\langle {{x_0}^2} \right\rangle }
\end{array}} \right]
\end{equation}

Therefore, the thermal emittance fitted using the equivalent transfer matrix $\tilde R$ should be 
\begin{equation}
{\varepsilon _{fit}} = \left| {{{\tilde \Sigma }_0}} \right| = 0 = {\varepsilon _0}
\end{equation}

The above analysis indicates that accurate thermal emittance can be fitted using the equivalent transfer matrix $\tilde R$. Next rigorous mathematical derivations are employed to verify the above theory. In a solenoid scan beamline (Fig.~\ref{Fig.sketch}) the initial beam matrix and initial emittance is expressed as the form shown in Eqn.~\ref{eq150}, thus the beam size squared at the screen $\sigma _{scr}^2$ can be expressed as the form shown in Eqn.~\ref{eq10}. The fitted beam moments at position 0 is assumed to be $\left\langle {{x_{fit}}^2} \right\rangle $, $\left\langle {{x_{fit}}{x_{fit}}^\prime } \right\rangle $, $\left\langle {{x_{fit}}{{^\prime }^2}} \right\rangle $. If $R$ is employed in the transfer matrix calculation the fitting routine would retrieve the initial real beam moments, i.e., $\left\langle {{x_{fit}}^2} \right\rangle  = \left\langle {{x_0}^2} \right\rangle $, $\left\langle {{x_{fit}}x'_{fit}} \right\rangle  = \left\langle {x'{{_{fit}}^2}} \right\rangle  = 0$. However, the fitted beam moments must not be equal to the real beam moments if $\tilde R$ is employed. The fitted beam sizes squared ${\sigma _{fit}}^2$ as a function of the fitted beam moments can be expressed as

\begin{equation}
{\sigma _{fit}}^2 = {{\tilde R}_{11}}^2\left\langle {{x_{fit}}^2} \right\rangle  + 2{{\tilde R}_{11}}{{\tilde R}_{12}}\left\langle {{x_{fit}}{x_{fit}}^\prime } \right\rangle  + {{\tilde R}_{12}}^2\left\langle {{x_{fit}}{{^\prime }^2}} \right\rangle 
\end{equation}

 Similar as the $\it Method~One$, we calculate ${\sigma _{scr}}^2 - {\sigma _{fit}}^2$ and factorize it in terms of $k_2$:

\begin{equation}
{\sigma _{scr}}^2 - {\sigma _{fit}}^2 = {p_0} + {p_1}{k_1} + {p_2}{k_2}^2 + {p_3}{k_2}^3 + {p_4}{k_2}^4
\end{equation}

The coefficient of the term ${k_2}^4$ is calculated to be

\begin{equation}
{p_4} = {\xi ^2}{L_1}^2{L_d}^2\left( {\left\langle {{x_0}^2} \right\rangle  - \left\langle {{x_{fit}}^2} \right\rangle } \right)
\end{equation}

$p_4=0$ when $\left\langle {{x_{fit}}^2} \right\rangle  = \left\langle {{x_0}^2} \right\rangle $. Therefore, $\left\langle {{x_{fit}}^2} \right\rangle $ is replaced by $ \left\langle {{x_0}^2} \right\rangle $ in the calculation of the coefficient of the term ${k_2}^3$:

\begin{equation}
{p_3} = 2\xi {L_1}^2{L_d}^2\left( {\left\langle {{x_{fit}}{x_{fit}}^\prime } \right\rangle  - {k_g}\left\langle {{x_0}^2} \right\rangle } \right)
\end{equation}

$p_3=0$ when $\left\langle {{x_{fit}}{x_{fit}}^\prime } \right\rangle  = {k_g}\left\langle {{x_0}^2} \right\rangle $. $\left\langle {{x_{fit}}{x_{fit}}^\prime } \right\rangle$ is replaced by $ {k_g}\left\langle {{x_0}^2} \right\rangle $ in the calculation of the coefficient of the term ${k_2}^2$:

\begin{equation}
{p_2} = {L_1}^2{L_d}^2\left( {{k_g}^2\left\langle {{x_0}^2} \right\rangle  - \left\langle {{x_{fit}}{{^\prime }^2}} \right\rangle } \right)
\end{equation}

$p_2=0$ when $\left\langle {{x_{fit}}{{^\prime }^2}} \right\rangle  = {k_g}^2\left\langle {{x_0}^2} \right\rangle  $. $\left\langle {{x_{fit}}{{^\prime }^2}} \right\rangle $ is replaced by ${k_g}^2\left\langle {{x_0}^2} \right\rangle $ in the calculation of the coefficients $p_1$ and $p_0$. It is found that $p_1=0$ and $p_0=0$.

As a result, there is certain $\left\langle {{x_{fit}}^2} \right\rangle $, $\left\langle {{x_{fit}}{x_{fit}}^\prime } \right\rangle $ and $\left\langle {{x_{fit}}{{^\prime }^2}} \right\rangle $ that make ${\sigma _{fit}}^2 = {\sigma _{scr}}^2$ under any solenoid strength $k_2$. The solutions of the equation ${\sigma _{scr}}^2 ={\sigma _{fit}}^2$ are

\begin{equation}\label{eq29}
\left\{ 
\begin{array}{l}
\left\langle {{x_{fit}}^2} \right\rangle  = \left\langle {{x_0}^2} \right\rangle \\
\left\langle {{x_{fit}}{x_{fit}}^\prime } \right\rangle  = {k_g}\left\langle {{x_0}^2} \right\rangle \\
\left\langle {{x_{fit}}{{^\prime }^2}} \right\rangle  = {k_g}^2\left\langle {{x_0}^2} \right\rangle 
\end{array}
\right.
\end{equation}

We found that the fitted beam moments are same as the moments in the equivalent initial beam matrix $\tilde \Sigma _0$ (Eqn.~\ref{eq2009}). Finally, using these fitted beam moments, we can calculate the measured (fitted) emittance at position 0 as
\begin{equation}
{\varepsilon _{fit}} = \sqrt {\left\langle {{x_{fit}}^2} \right\rangle \left\langle {x{{_{fit}'}^2}} \right\rangle  - {{\left\langle {{x_{fit}}x_{fit}'} \right\rangle }^2}} =0 = {\varepsilon _0}
\end{equation}

The fitted emittance at position 0 is equal to the actual initial emittance.

It should be noted that the new method is error-free based on a premise that the rf ($-k_g$) and solenoid ($k_{s1}$) fields overlap in the thin-lens approximation. If $k_{s1}$ is before $-k_g$ and  there is a distance $\delta$ between $k_{s1}$ and $-k_g$, i.e., the transfer matrix from the cathode to the screen is
\begin{equation}
\begin{array}{l}
R_2 = \left[ {\begin{array}{*{20}{c}}
1&{{L_d}}\\
0&1
\end{array}} \right]\left[ {\begin{array}{*{20}{c}}
1&0\\
{ - {k_2}}&1
\end{array}} \right]\left[ {\begin{array}{*{20}{c}}
1&{{L_1}}\\
0&1
\end{array}} \right]\left[ {\begin{array}{*{20}{c}}
1&0\\
{{k_g}}&1
\end{array}} \right] \times \\
\;\;\;\;\;\;\;\;\left[ {\begin{array}{*{20}{c}}
1&\delta \\
0&1
\end{array}} \right]\left[ {\begin{array}{*{20}{c}}
1&0\\
{ - {k_1}}&1
\end{array}} \right]
\end{array},
\end{equation}the new method is not strictly error-free. However, the measurement error of the thermal emittance due to the distance $\delta$ is negligible, which will be verified in the following simulations, because $\delta  \ll {L_1} + {L_d}$ is satisfied for almost all solenoid-scan beamlines.

\section{simulations}\label{section6}

The above analytic results for the solenoid scan are verified with numerical simulations using ASTRA but now will include a nonzero initial emittance. In the simulation results below, the cathode gradient is set to 32 MV/m, and the laser launch phase with respect to rf is set to $35^{\circ}$. The space charge effect is excluded in the simulation. The initial electron beam spot size on the cathode has a uniform transverse distribution with 2 mm diameter (rms spot size 0.5 mm). The thermal emittance is assumed to be 1.0956 mm mrad/mm. Therefore, the estimated rms thermal emittance is 0.5 mm $\times$ 1.0956 mmmrad/mm, or 0.5478 mm mrad. The initial electron beam has a Gaussian longitudinal distribution with 100 fs FWHM pulse length, and such a short pulse length makes the emittance growth due to the phase-dependent transverse kick in the rf gun negligible.

The gun and solenoid field profiles from three groups, THU, AWA and PITZ, as exhibited in Fig.~\ref{Fig.field_plot}, are employed in the simulations. A screen is placed at 3 m downstream from the cathode. In the ASTRA simulation of the solenoid scan measurement, the solenoid strength is scanned and a series of spot sizes at the screen are generated. The position that divides the overlapping and non-overlapping solenoid fields $L_{sep}$ are calculated based on the field profiles, as shown in Table 1, so as the strength ratio of the two parts of the solenoid $\xi$. $\xi$ indicates the overlap severity of the rf and solenoid fields. In the three beamlines from THU, AWA and PITZ, $\xi$ of the PITZ beamline is the largest while $\xi$ of the THU beamline is the smallest.

\begin{table*}[hbtp]
\caption{\label{t12} Comparing the fitting results of three solenoid scan beam lines. $\epsilon_{th}$ is the thermal emittance,  $L_{sep}$  is the position that divides the overlapping and non-overlapping solenoid fields. $\xi$ is the strength ratio of the overlapping and non-overlapping solenoid fields. ${\epsilon_{fit,1}}$ is the fitting result based on the $\it Method~One$. $\tilde\epsilon_{fit,new} $ is the fitting result of the new method assuming $p_z(z)=p_f$. $\epsilon_{fit,new} $ is the fitting result of the new method using $p_z(z)  $ abtained from ASTRA simulation. The unit of the emittances is mm~mrad.}
\renewcommand\tabcolsep{8pt}
\renewcommand\arraystretch{1.5}
\begin{tabular}{*{10}{c}}
\toprule[1.5pt]
  beamline       &  $L_{sep}$ [m]   & $\xi$ &$\epsilon_{th}$ & ${\epsilon_{fit,1}}$  &$\frac{{\left| {{\epsilon _{fit,1}} - {\epsilon _{th}}} \right|}}{{{\epsilon _{th}}}}$  &  $\tilde\epsilon_{fit,new} $  &$\frac{{\left| {{\tilde\epsilon _{fit,new}} - {\epsilon _{th}}} \right|}}{{{\epsilon _{th}}}}$  &  $\epsilon_{fit,new} $  &$\frac{{\left| {{\epsilon _{fit,new}} - {\epsilon _{th}}} \right|}}{{{\epsilon _{th}}}}$   \\
\midrule[1pt]
THU             &0.1062 &0.0363 & 0.5478&0.5488 &$0.18\%$ &  0.5471&$0.13\%$ &0.5471&0.13\% \\ 
AWA          &0.2162 & 0.2103& 0.5478  &0.5874 &$7.2\%$ & 0.5477 & $0.02\%$&0.5477&0.02\%\\ 
PITZ            &0.2338& 0.5349& 0.5478&0.6917 &$26.3\%$ & 0.5596 & $2.2\%$&0.5472&0.11\%\\ 
\bottomrule[1.5pt]
\end{tabular}
\end{table*}

Firstly, the beam spot sizes versus the solenoid strength are fitted to calculate the emittance based on the $\it Method~One$. The fitted emittances (${\epsilon_{fit,1}}$) are listed in Table 1. The overestimation of the thermal emittance is $0.18\%$, $7.2\%$, $26.3\%$ for the THU, AWA, and PITZ beamlines respectively. The overestimation is larger for a larger $\xi$. The measurement error of the thermal emittance is acceptable for the THU beamline, but too large for the PITZ beamline.

Secondly, the thermal emittance is calculated based on the new method. It should be noted that  theoretical analysis of the new method demonstrated in Sec.~\ref{section5} uses a simplified model, in which the overlapping solenoid is simplified to a thin lens $k_{s1}$ located at the gun exit. Therefore, the fitted beam moments (Eqn.~\ref{eq29}) are at the gun exit. In the simulation we calculate the transfer matrix from the cathode to the screen to ensure that the complete overlapping solenoid field is considered, which is equivalent to adding a drift section before position 0 of the simplified model (Fig.~\ref{Fig.sketch}), and will not change the thermal emittance fitting result due to the weak focusing approximation. Therefore, the transfer matrix calculation includes both the overlapping and non-overlapping solenoid fields, the drift, but not the rf field. The equivalent field of the overlapping solenoid $B'_z(z)$ (Eqn.~\ref{eq1009}) should be used in the transfer matrix calculation.  $B'_z(z)$ contains the term of $p_z(z)$, which is determined by the axial rf field. Here $p_z(z)$  is obtained by importing the axial rf field profile into ASTRA simulation. The beam spot sizes versus the solenoid strength are fitted to calculate the emittance, and the fitted emittances (${\epsilon_{fit,new}}$) are listed in Table 1. The fitting error of the thermal emittance is negligible ($<0.13\%$) for all the THU, AWA, and PITZ beamlines, proving that the measurement error of the thermal emittance with the new method is negligible. 

It seems that $p_z(z)$ in the new method requires accurate knowledge of the axial rf field profile, which is similar to the case of the $\it Method ~Two$. However, as shown in Fig.~\ref{Fig.field_plot}, $B_z(z)$ of the overlapping solenoid increases along $z$ axis from the cathode to $L_{sep}$. The interval with strong $B_z(z)$ is mainly located around the gun exit. $B_z(z)$ of the overlapping solenoid is small in the interval of beam acceleration, thus the influence of the uncertainty of $p_z(z)$ on the thermal emittance fitting result should be small. For example, we make a rough approximation of $p_z(z)$ that $p_z(z)=p_f$, i.e., $B'_z(z)=B_z(z)$, and the emittance are calculated based on the new method. The fitted emittance (${\tilde\epsilon_{fit,new}}$) are listed in Table I. The error of the thermal emittance is $0.13\%$, $0.02\%$, $2.2\%$ for the THU, AWA, and PITZ beamlines respectively. The measurement errors of the THU and AWA beamlines are negligible. Even the measurement error of the PITZ beamline is acceptable for most occasions that do not require extreme measurement accuracy. 

\begin{figure}[hbtp]
	\centering
	\includegraphics[scale=0.75]{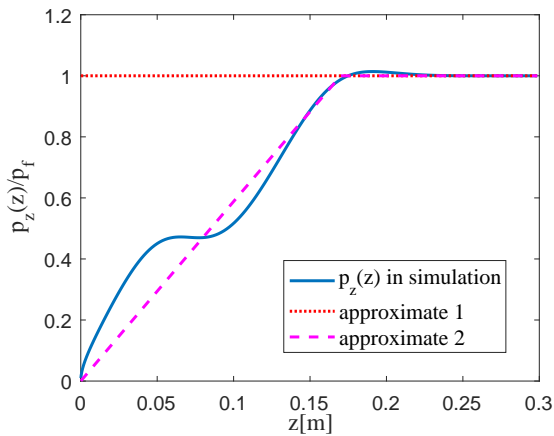}
	\caption{\label{Fig.pz_plot}  blue line: $p_z(z)/p_f$ simulated in ASTRA using the axial rf field profile; red dots: approximation 1 of $p_z(z)$, $p_z(z)=p_f$; magenta dashed line: approximation 2 of $p_z(z)$ (see Eqn.~\ref{eq1010}).}
\end{figure}

We further study the measurement error of the PITZ beamline using a more reasonable approximation of $p_z(z)$. As shown in Fig.~\ref{Fig.pz_plot}, the blue line is $p_z(z)/p_f$ simulated in ASTRA using the real axial rf field profile; the red dots are the rough approximation of $p_z(z)$: $p_z(z)=p_f$; the magenta dashed line is a new approximation of $p_z(z)$:
\begin{equation}\label{eq1010}
{p_z}(z) =  \left\{{\begin{array}{*{20}{c}}
{{p_f}\times z/0.17 ~\rm m ,\;z < 0.17 ~\rm m}\\
{{p_f},~~~~~~~~~~~~~~~~~~\;z >  = 0.17 ~\rm m}
\end{array}} \right.
\end{equation}

With the new approximation of $p_z(z)$ demonstrated in Eqn.~\ref{eq1010}, the fitted emittance of the PITZ beamline is 0.5483 mm mrad.  The error of the thermal emittance is only $0.09\%$, which is negligible. From this we can conclude that the accuracy of $p_z(z)$ has little effect on the fitting result of the thermal emittance with the new method.

\section{experiments}\label{section7}

\begin{figure}[hbtp]
	\centering
	\includegraphics[scale=0.75]{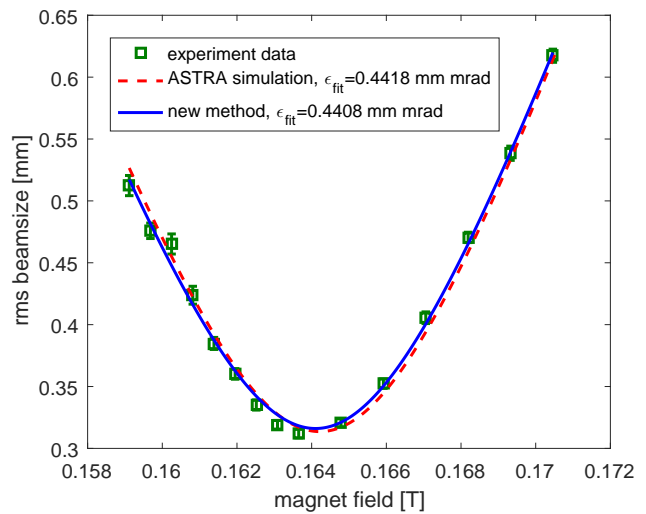}
	\caption{\label{Fig.desy_experiment} green squares: experiment data of the beam spot sizes versus the solenoid strength in the PITZ beamline. Red dashed line: fitting curve based on the $\it Method~Two$. Blue solid line: fitting curve based on the new method.}
\end{figure}

Two experiments of solenoid scans from PITZ and AWA respectively are demonstrated to further verify the performance of the new method. For the solenoid scan in the PITZ beamline, the cathode gradient was 53 MV/m and the laser launch phase with respect to rf was $42^\circ$ resulting in beam energy of 5.336~MeV.  The screen was placed at 5.28 m downstream from the cathode. The laser rms spot size was 0.38 mm. The measured beam spot sizes versus the solenoid strength are illustrated in Fig.~\ref{Fig.desy_experiment}. The rms beam size was calculated as the geometric average of the sizes in the $x$ and $y$ directions: $\sigma=\sqrt{\sigma_{x}\sigma_{y}}$. Firstly, the thermal emittance was fitted based on the $\it Method~Two$. The initial emittance on the cathode was scanned in ASTRA simulation until the measured beam spot sizes best agree with the simulation. The best-fitting result is shown in the red dashed line in Fig.~\ref{Fig.desy_experiment}, and the fitted emittance is 0.4418 mm mrad. The fitted curve is in good agreement with the measurements, indicating that the rf field profile in the simulation is very close to the real profile of the PITZ gun. Therefore, $\it Method~Two$ is feasible to be used to measure the thermal emittance in the PITZ beamline, and the fitted emittance should be reliable. On the other hand, the thermal emittance is fitted based on the new method. In the new method $p_z(z)$ is obtained in ASTRA simulation using the rf field profile to calculate $B'_z(z)$, and the fitted emittance is 0.4408 mm mrad, which is very close to the measured emittance using the $\it Method~Two$, proving that the new method is accurate in thermal emittance measurements with solenoid scans.

\begin{figure}[hbtp]
	\centering
	\includegraphics[scale=0.8]{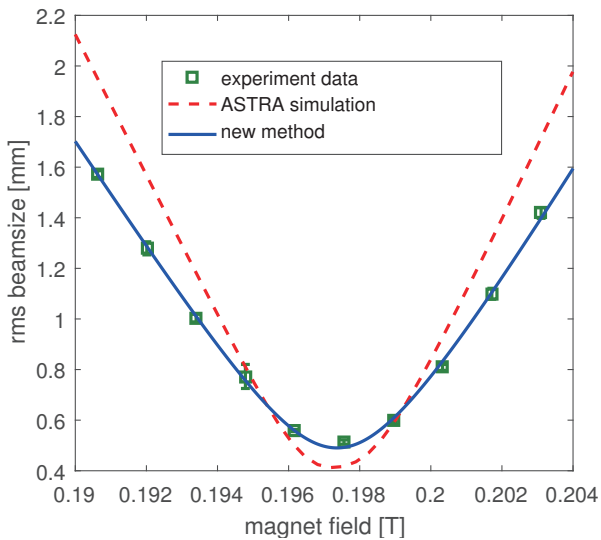}
	\caption{\label{Fig.awa_experiment}  green squares: experiment data of the beam spot sizes versus the solenoid strength in the AWA beamline. Blue solid line: fitting curve based on the new method. Red dashed line: simulated results in ASTRA using the thermal emittance fitted with the new method.  }
\end{figure}

Moreover, the beam spot sizes versus the solenoid strength measured in the AWA beamline are illustrated in Fig.~\ref{Fig.awa_experiment}. The rms beam size was also calculated as the geometric average of the sizes in the $x$ and $y$ directions. In the experiment the cathode gradient was 32 MV/m and the laser launch phase with respect to rf was $43^\circ$ resulting in beam energy of 3.2~MeV.  The screen was placed at 2.98 m downstream from the cathode. The laser rms spot size was 2.7 mm. Firstly, the thermal emittance was fitted based on the $\it Method~Two$. The initial emittance on the cathode was scanned in ASTRA simulation to fit the measured data. The electron emission on the cathode is isotropic in the simulation based on the three-step model \cite{dowell2009quantum}, i.e., $\left\langle {{x_{0}}{x_{0}}^\prime } \right\rangle =0$. We found that it is impossible to have a good fitting for any initial emittance on the cathode, as shown in the red dashed line in Fig.~\ref{Fig.awa_experiment}, which indicates that there is a discrepancy between the simulated and real rf field profile of the AWA gun, or the relative distance between the solenoid and the rf gun used in the simulation is different from the real beamline. Of course we can set a $\left\langle {{x_{0}}{x_{0}}^\prime } \right\rangle $ on the cathode to achieve a good agreement between the simulations and measurements, but it's not consistent with the basic physics model. In summary, the $\it Method~two$ has a limitation that it can't be used when the rf field profile or the distance between the elements is not accurately knowledged. Secondly, the thermal emittance is fitted based on the new method using the approximation of $p_z(z)=p_f$, shown as the blue line in Fig.~\ref{Fig.awa_experiment}. Simulation in Sec.~\ref{section6} shows that the measurement error is negligible with this approximation. Some solenoid scan experiments done in the AWA beamline have used the new method introduced in this paper to fit the thermal emittance. The measured thermal emittance of a cesium telluride photocathode is $1.05\pm0.04$ mm mrad/mm, which is in good agreement with the theoretical value \cite{zheng2018overestimation,PhysRevAccelBeams.22.072805}, proving that the new method is appropriate to be employed in solenoid scans when the rf field profile or the distance between the elements is not accurately knowledged.

\section{conclusion}
The measurement uncertainty of the thermal emittance due to the rf and solenoid fields overlap in solenoid scans has been systematically studied in this paper. Two conventionally used methods to solve the overlap issue are summarized: $\it Method~One$ uses the non-overlapping part of the solenoid in the thermal emittance fitting and abandons the overlapping part. This method leads to an overestimation of the thermal emittance, which has been proved by theoretical derivations. $\it Method~Two$ uses the beam dynamics simulation from the cathode to the screen including the rf field in the thermal emittance fitting. The method can't be used if the rf field profile or the distance between the elements is not accurately knowledged.

A new method is provided in this paper to solve the overlap issue. The transfer matrix from the cathode to the screen is calculated including the complete solenoid but excluding the rf field. The magnetic field of the overlapping solenoid is replaced by an equivalent field containing a term of $p_z(z)$. Theoretical derivations and ASTRA simulations in three different beamlines (THU, AWA and PITZ) demonstrate that the fitted emittance is equal to the thermal emittance, i.e., the measurement error of the thermal emittance with the new method is negligible. Further study also shows that the accuracy of $p_z(z)$ has little effect on the fitting result of the thermal emittance with the new method. Even under a rough approximation of $p_z(z)=p_f$, the fitting error is negligible ($<0.13\%$) for the THU and AWA beamlines, and acceptable ($<2.2\%$) for the PITZ beamline. Finally, experiment data of solenoid scans from PITZ and AWA are processed using the new method, further verifying that the measurement error with the new method is negligible.

Since most of the normal-conducting rf photoinjectors have the issue of the rf and solenoid fields overlap, we believe that our new method is a fundamental complement to the solenoid scan technique, and the application of this method will significantly improve the accuracy of thermal emittance measurements.

\section*{Acknowledgments}
This work is supported by the National Natural Science Foundation of China (NSFC) under Grant No. 11435015 and No. 11375097. It is also funded by the Postdoctoral Science Foundation of China under Grant No. 2019M660670. Besides, we thank Dr. Houjun Qian in DESY for the productive discussions about this paper.

\bibliography{mybibfile.bib}

\end{document}